\documentclass{iau}

\usepackage{amsmath}
\usepackage{graphicx}
\usepackage{multirow}
\usepackage{xcolor}

\newcommand{\rb}{$\langle R_{\rm b} \rangle$}
\newcommand{\rg}{$\langle R_{\rm g} \rangle$}
\newcommand{\amax}{$a_{\rm max}$}

\def\apj{ApJ}

\def\apjs{ApJS}

\def\aap{A\&A}

\def\mnras{MNRAS}

\defcitealias{Dantas2025a}{Paper I}
\defcitealias{Dantas2025b}{Paper II}
\defcitealias{GarciaDelgado2026}{Paper III}

\begin{document}

\lefttitle{M.~L.~L.~Dantas et al.}
\righttitle{}

\jnlPage{1}{7}
\jnlDoiYr{2026}
\doival{10.1017/xxxxx}

\aopheadtitle{Proceedings IAU Symposium}
\editors{G. Buldgen,  
         A. Vidotto \& 
         A. Miglio, eds.}

\title{Galactic archaeology meets exoplanets: \\ linking stellar birth radii to exoplanet demographics}


\author{M. L. L. Dantas$^{1}$, 
       J. J. García-Delgado$^{2}$, 
       I. Rebollido$^{3}$, 
       R. Smiljanic$^{4}$
       }

\affiliation{
        $^{1}$Instituto de Astrofísica, Pontificia Universidad Católica de Chile, Av. Vicuña Mackenna 4860, Santiago, Chile\\
        $^{2}$Space Research Group, Universidad de Alcalá, 28805 Alcalá de Henares, Spain\\
        $^{3}$ Centro de Astrobiolog\'ia (CAB) CSIC-INTA, Camino Viejo del Castillo s/n, E-28692, Villanueva de la Cañada, Madrid, Spain\\
        $^{4}$ Nicolaus Copernicus Astronomical Center, Polish Academy of Sciences, ul. Bartycka 18, 00-716, Warsaw, Poland}

\begin{abstract}
Stars observed in the solar vicinity were not necessarily born there. Radial mixing can redistribute stellar populations across the Galactic disc, decoupling their present-day locations from the environments in which they formed. We summarise a framework that combines Galactic chemical evolution models with a generalised additive model to infer stellar birth radii. We apply it to 1327 predominantly thin-disc planet hosts with homogeneous ages, \textit{Gaia} DR3 astrometry, and \texttt{SWEET-Cat} atmospheric parameters. We find that giant-planet systems preferentially trace metal-rich inner-disc birth environments, whereas rocky-only systems are less centrally concentrated; and no clear relation between radial displacement and detected planet multiplicity. Planet-host stars that migrate towards the inner parts of the Galaxy have systems with larger orbital separations; however, this may partly reflect the known dependence of planetary orbital properties on host-star metallicity rather than a causal effect of Galactic stellar migration. These results illustrate how Galactic archaeology can complement exoplanet demographics by linking planetary systems to their probable birth environments.
\end{abstract}


\maketitle

\section{Introduction}

The Milky Way disc is not static. Stars interact gravitationally with the Galactic bar, spiral structure, giant molecular clouds, and other perturbations, which can progressively alter their orbits. Radial mixing is commonly separated into \emph{blurring}, in which stars undergo epicyclic excursions without a net change in angular momentum, and \emph{churning}, or radial migration, in which their guiding radii change \citep[e.g.][]{SellwoodBinney2002}. As a consequence, stars presently sharing the solar vicinity may have formed under substantially different chemical and dynamical conditions.

This line of work was motivated by \citet{Dantas2023}, where we identified old super-metal-rich stars in the solar vicinity whose chemical properties were more compatible with formation in the inner Galactic disc than with their present-day locations. This led us to set up a method capable of estimating stellar birth radii directly. In \citet{Dantas2025a}, we developed a semi-empirical framework that combines the time-dependent thin-disc chemical-evolution models of \citet{Magrini2009} with a generalised additive model (GAM). Applied to \emph{Gaia}--ESO thin-disc stars, the method revealed the expected chemo-dynamical pattern: metal-rich stars formed preferentially in the inner disc and were predominantly displaced outwards, while metal-poor stars formed at larger radii and more frequently moved inwards.

This framework subsequently developed into the \emph{Probing the origins} series. \citetalias{Dantas2025a} introduced the GAM-based inference of Galactic birth radii \citep{Dantas2025a}, while \citetalias{Dantas2025b} used the resulting estimates to test whether radial migration leaves an imprint on stellar Li depletion \citep{Dantas2025b}, with the Solar Li abundance considered separately by \citet{Dantas2025c}. A particularly illustrative result from \citetalias{Dantas2025a} is the inferred Solar birth radius of $7.08\pm0.24$ kpc, with a $3\sigma$ interval of 6.46--7.81 kpc, interior to its present Galactocentric location. If the Solar System itself has travelled across the Galactic disc, this naturally raises a broader question: \emph{do planetary systems retain observable information about the Galactic environments in which they formed?}

\citetalias{GarciaDelgado2026}  addresses this question by applying the birth-radius framework to known exoplanetary systems \citep{GarciaDelgado2026}. This proceedings contribution summarises the combined application of Papers I and III, with particular emphasis on the connection between Galactic birth environments, stellar radial displacement, and present-day exoplanet demographics.

\section{The planet-host sample}

We constructed a working sample of planet-host stars from confirmed systems listed in the \textit{Encyclopaedia of Exoplanetary Systems} \citep{Schneider2011}. The host stars were cross-matched with \textit{Gaia} DR3 \citep{GaiaDR3} for astrometry and radial velocities and with \texttt{SWEET-Cat} \citep{Santos2013} for homogeneous effective temperatures ($T_{\rm eff}$) and metallicities ([Fe/H]). Astrometric quality criteria based on \textit{Gaia} diagnostics were applied to reduce contamination by problematic solutions and unresolved multiplicity. 

An important limitation is that the known exoplanet census samples the Galactic disc very non-uniformly, with strong spatial concentrations associated with specific surveys and fields (e.g. the \textit{Kepler} mission). Our demographic trends should therefore be interpreted in light of this spatial selection, the heterogeneous detection methods, and the incompleteness of individual planetary systems. These limitations should progressively lessen as new data releases and surveys (including \textit{Gaia} DR4, PLATO, Roman, and Rubin/LSST, to mention a few) expand the volume, precision, and homogeneity of the available stellar and planetary samples.

Homogeneous stellar ages were derived with version 1.5.3 of the BAyesian STellar Algorithm (\textsc{Basta}; \citealt{AguirreBorsenKoch2022}), using the solar-scaled BaSTI 2018 isochrone grid \citep{Hidalgo2018}. The fits combine \texttt{SWEET-Cat} $T_{\rm eff}$ and [Fe/H] with \textit{Gaia} DR3 parallaxes and $G$-band photometry. We adopt the median of the marginalised posterior age distribution and its 16th and 84th percentiles as the corresponding uncertainties. Successful homogeneous ages were obtained for 1327 hosts, which define the final working sample.

Galactic orbits were integrated with \textsc{Galpy} \citep{Bovy2015} in the Milky Way potential of \citet{McMillan2017}. Observational uncertainties were propagated by resampling the input astrometry and kinematics, yielding distributions of guiding radius, eccentricity, maximum vertical excursion, and orbital actions. The sample is dominated by FGK-type stars, with a small number of A-type interlopers belonging to the Galactic thin disc; also, since there is an overlap between the thin and thick discs, we retained the intermediate population as well in our analysis. Because the chemical evolution models used below describe the thin disc, likely thick-disc and halo members were excluded.

\section{Inferring Galactic birth radii with a generalised additive model}

The fundamental difficulty in recovering a stellar birth radius is that the present-day metallicity gradient does not describe the Galactic disc at the epoch when an old star formed. The chemical-evolution calculations of \citet{Magrini2009} provide radial abundance gradients at several evolutionary times, including predictions for [Fe/H] at 2.1, 3.3, 8.0, 11.0, and 13.7 Gyr (i.e. the ages of the Universe at star formation). Their discrete temporal and radial sampling, however, limits direct inference for individual stars.

The GAM presented in \citet{Dantas2025a} provides a flexible interpolation of this age--metallicity--radius relation. In practice, the smooth chemical-evolution predictions are used to construct a continuous mapping between stellar age, [Fe/H], and Galactic radius. For each observed star, uncertainties in age and metallicity are propagated through repeated sampling, producing a distribution of predicted birth radii. We use its median as \rb\ and the corresponding percentile interval as its uncertainty.

\begin{figure}[!t]
    \centering
    \includegraphics[width=0.49\linewidth]{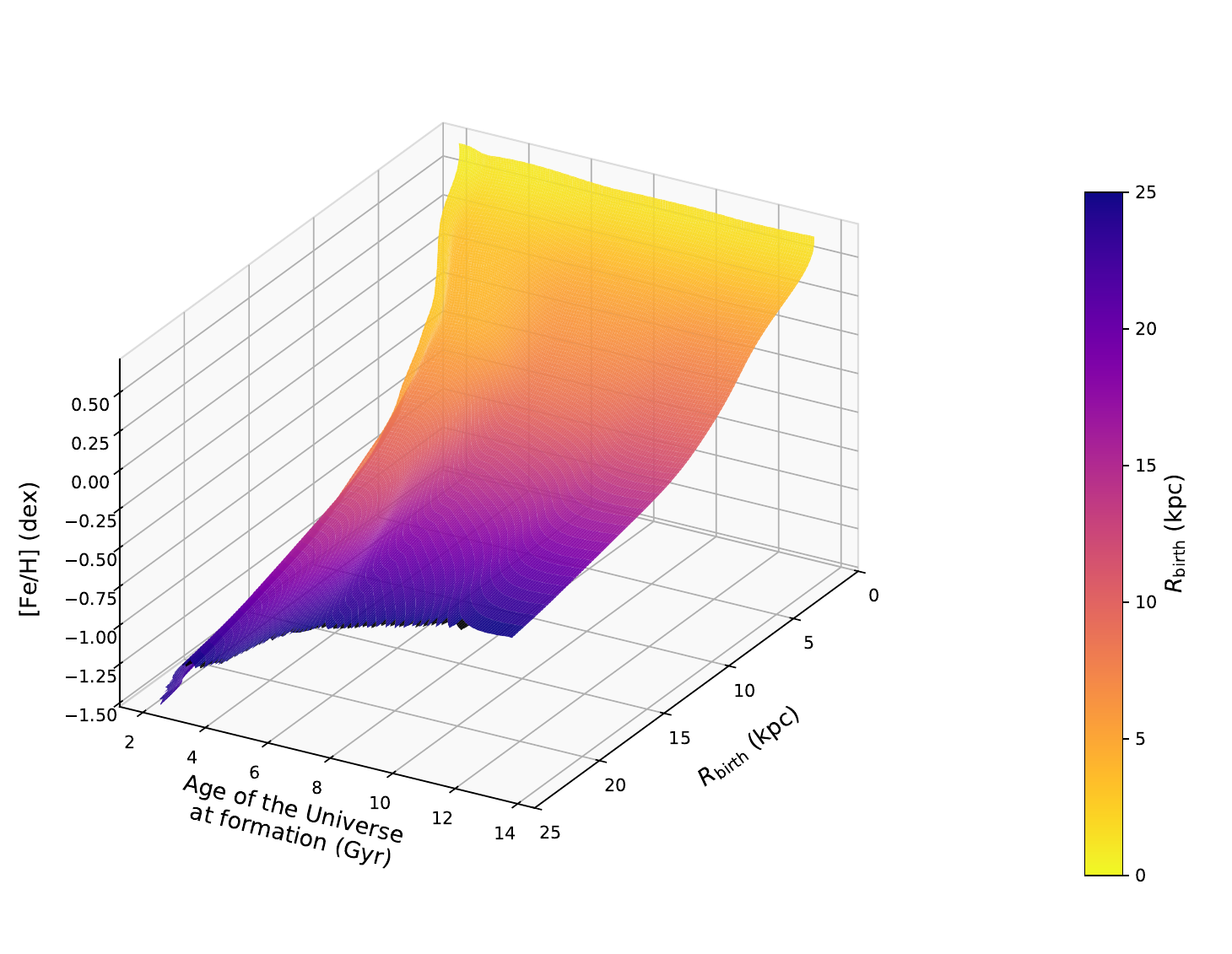}
    \includegraphics[width=0.49\linewidth]{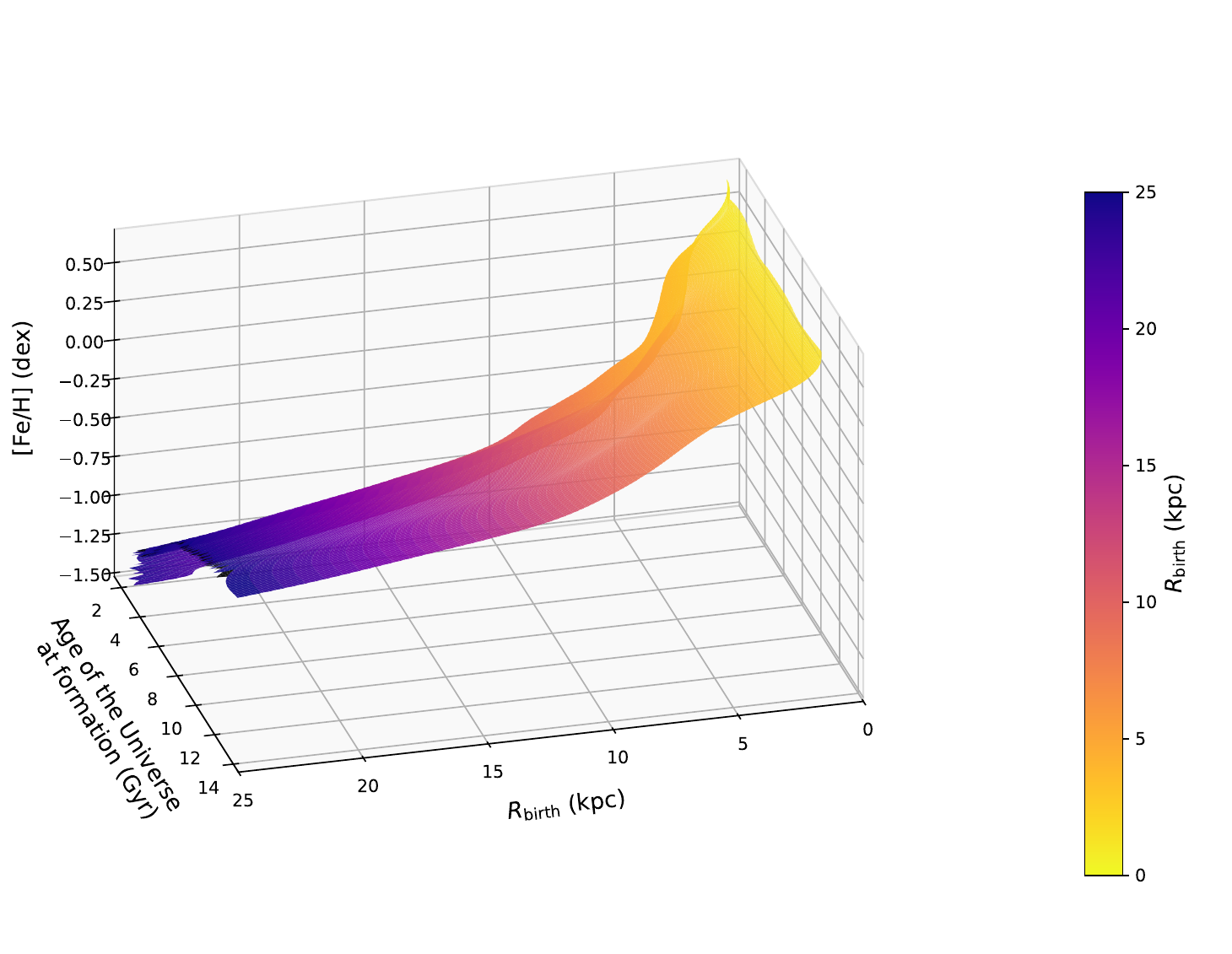}
    \includegraphics[width=0.49\linewidth]{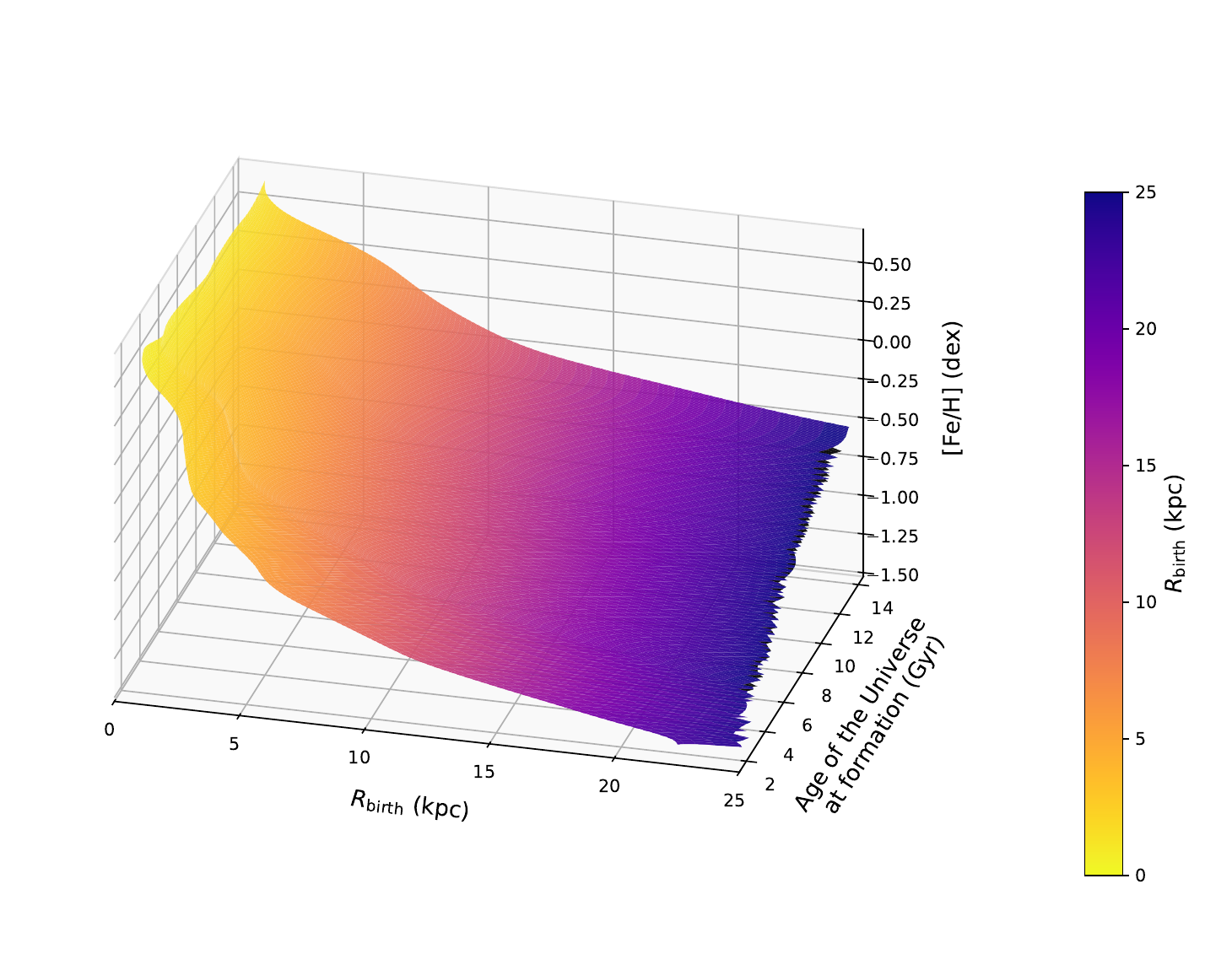}
    \includegraphics[width=0.49\linewidth]{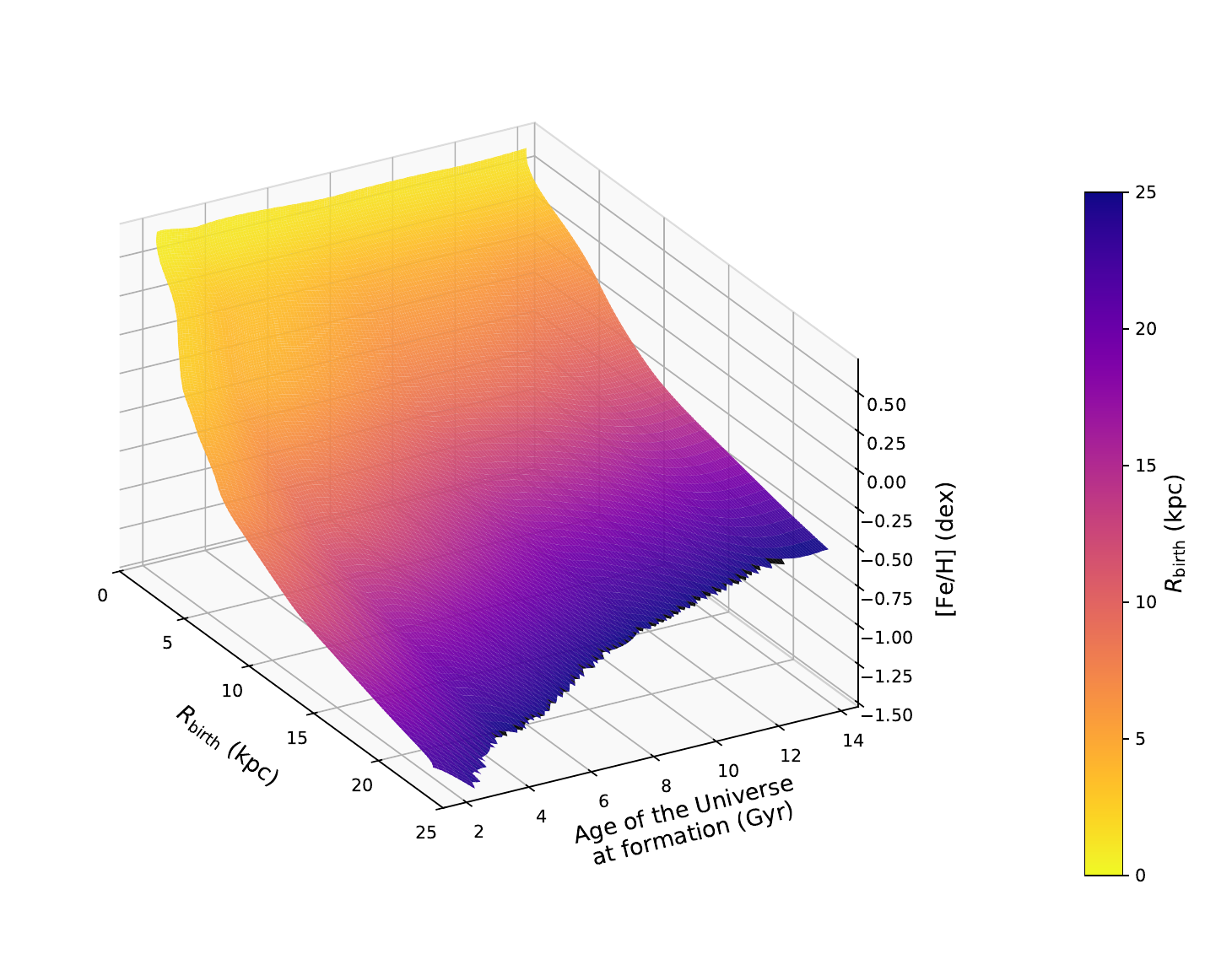}
    \caption{Four viewing angles of the GAM representation used for the birth-radius inference. The model continuously interpolates the time-dependent [Fe/H] gradients of \citet{Magrini2009} in age--metallicity--radius space, allowing characteristic birth radii to be inferred from measured ages and metallicities. Colour encodes the predicted birth radius.}
    \label{fig:gam}
\end{figure}

Figure~\ref{fig:gam} visualises this procedure in three dimensions. Such a representation is particularly useful because the inferred \rb\ is not a function of metallicity alone: the same [Fe/H] can correspond to different formation radii at different epochs. The recovered radius should therefore be interpreted as the location favoured by the adopted smooth thin-disc chemical-evolution model, rather than as a direct reconstruction of a star's past trajectory. In particular, intrinsic scatter of the interstellar medium around the mean radial gradient is not explicitly encoded in the present implementation.

We then compare \rb\ with the median guiding radius \rg. Stars for which the two quantities agree within $2\sigma$ are classified as non-migrators for the purposes of this analysis. Among the remaining systems, those with \rg$>$\rb\ are classified as having experienced a net outward displacement, while those with \rg$<$\rb\ are classified as having experienced a net inward displacement. These labels describe the inferred net radial displacement and should not be interpreted as a complete dynamical history of an individual star.

\section{Exoplanet demographics across Galactic birth environments}

Applying this framework to planet-hosting stars reveals that the exoplanet population currently observed around the solar vicinity samples a considerably broader range of Galactic formation environments than its present spatial distribution alone would suggest. Most hosts are inferred to have formed at smaller Galactocentric radii than their present guiding radii, while a smaller population is associated with formation in the outer disc. This broadly mirrors the radial mixing behaviour found for the \emph{Gaia}--ESO field-star sample in \citet{Dantas2025a}.

The inferred birth locations also vary with planetary architecture. Systems containing giant planets preferentially occupy smaller \rb, consistent with the strong connection between giant-planet occurrence and stellar metallicity. Rocky-only systems have larger characteristic birth radii and smaller collective radial displacements, whereas intermediate categories (i.e. rocky+giants) are formed in intermediate radii \citep{GarciaDelgado2026}. By contrast, we find no clear systematic relation between inferred radial displacement and the number of detected planets. This is important because it suggests that the strongest signal is not simply one of system multiplicity.

\begin{figure}[!t]
    \centering
    \includegraphics[width=\linewidth]{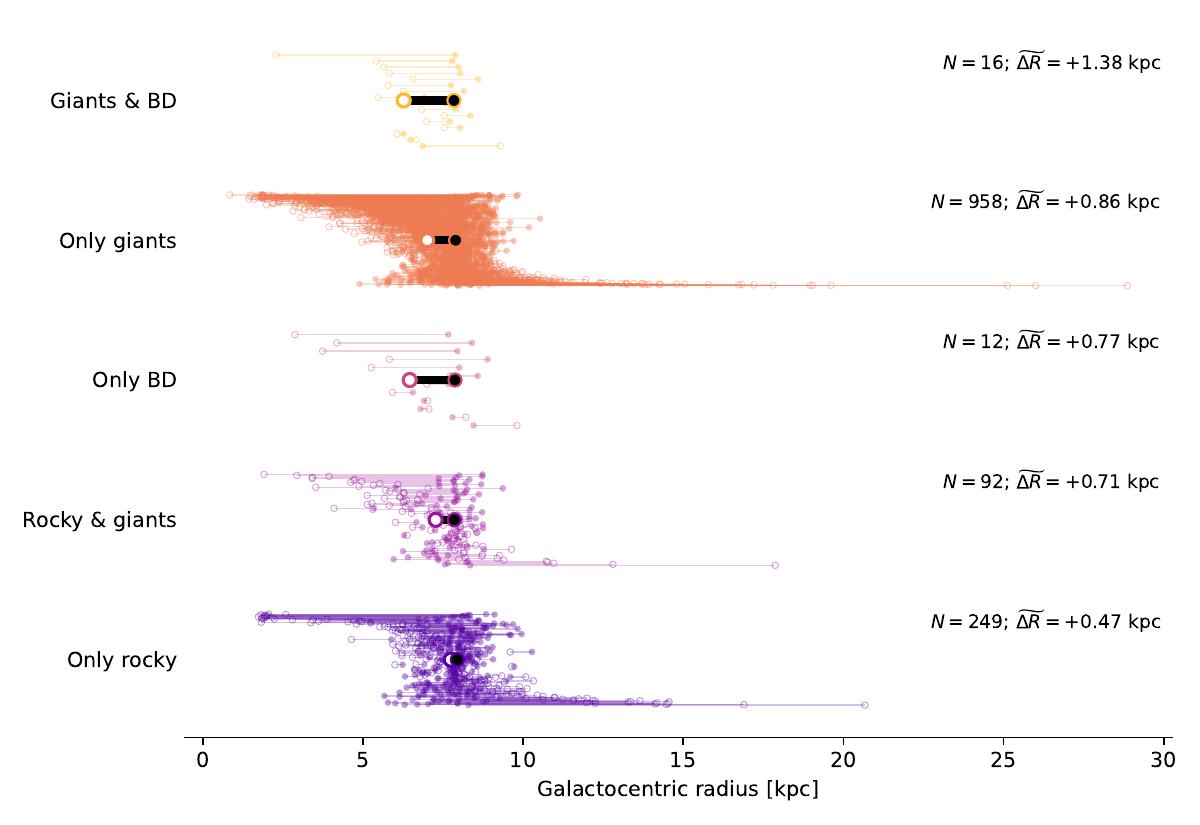}
    \caption{Radial displacement of planet-hosting systems grouped by planetary architecture. For each host, a horizontal segment connects the median inferred birth radius, \rb, to the median guiding radius, \rg. Open and filled markers indicate \rb\ and \rg, respectively, while the larger markers and thicker segment show the corresponding median values for each planetary category. Systems are vertically ordered by $\Delta R=\langle R_{\rm g}\rangle-\langle R_{\rm b}\rangle$ within each category. The annotations report the number of systems and median $\Delta R$ in each group.}
    \label{fig:demographics}
\end{figure}

Figure~\ref{fig:demographics} visualises these differences by directly connecting each host's inferred \rb\ to its present-day \rg. Rocky-only systems show the smallest median displacement, $\widetilde{\Delta R}=+0.47$ kpc, while the larger values among brown-dwarf-containing systems should be interpreted cautiously given their small sample sizes.

\begin{figure}[!t]
    \centering
    \includegraphics[width=\linewidth]{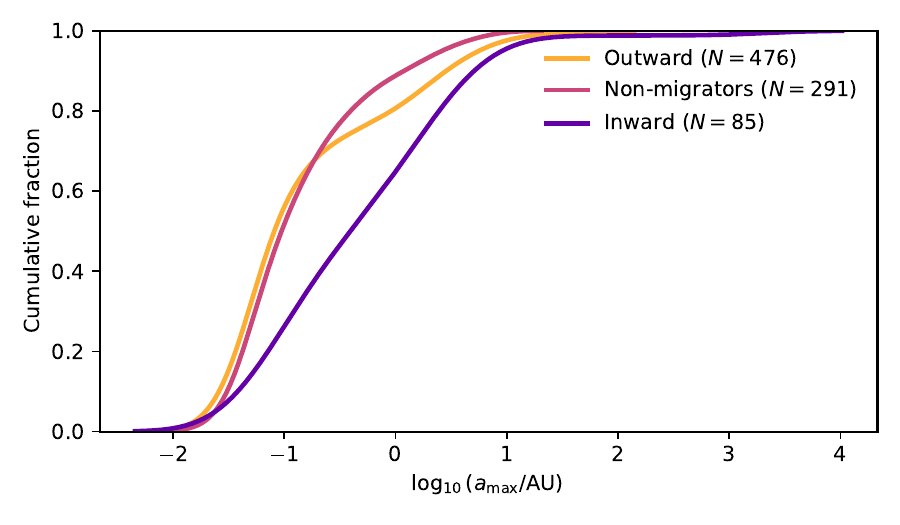}
    \caption{Smoothed cumulative distributions of \amax\ for outward- non-, and inward-migrators, obtained by integrating Gaussian kernel-density estimates in $\log_{10}(a_{\rm max}/{\rm AU})$. Inward-moving systems extend towards larger \amax, while the other two populations overlap substantially. Sample sizes are given in the legend.}
    \label{fig:amax}
\end{figure}

A further difference emerges when considering the semi-major axis of the outermost detected companion (\amax). Figure~\ref{fig:amax} shows the smoothed cumulative distributions of \amax\ for the three radial-displacement classes. Inward-moving hosts are shifted towards larger \amax, whereas outward-moving systems and non-migrators overlap substantially. This should not be interpreted as evidence that stellar radial migration directly restructures planetary systems: planets around metal-poor stars are known to preferentially occupy longer-period orbits \citep{Adibekyan2013}, and metallicity is itself intrinsically connected to the birth-radius inference used here. The \amax\ trend may therefore reflect, at least partly, differences in host-star metallicity rather than a causal effect of Galactic motion. Interpretation is further complicated by heterogeneous detection methods, spatial selection effects, and incomplete planetary-system censuses.

A less speculative interpretation is nevertheless astrophysically useful: stars that formed at different Galactic radii experienced different chemical environments, and the present-day exoplanet population retains measurable demographic differences between those environments. Galactic archaeology can thus complement the usual description of planetary systems in terms of host mass, metallicity, and age by adding another dimension: \emph{where in the Galaxy the system most probably formed}. Disentangling the effects of birth environment, metallicity, stellar radial displacement, and observational selection will require more homogeneous planetary samples.

Our results are qualitatively consistent with \citet{Teixeira2025}, who inferred Galactic birth radii from stellar age and [Fe/H], following the framework by \citet{Minchev2018}, and found that hosts of high-mass planets tend to originate at smaller Galactocentric radii than hosts of low-mass planets. While their analysis focused on birth radii alone, the present work additionally compares them with guiding radii and Galactic orbital properties, allowing the connection between planetary demographics and stellar radial displacement to be explored.

More broadly, the interface between Galactic archaeology and exoplanet science is rapidly becoming a community effort. Observational studies, as well as simulations, have begun to connect planet occurrence and architecture with the chemistry, ages, kinematics, and Galactic populations of their host stars, while complementary theoretical work explores how planet demographics vary across the evolving Milky Way \citep[e.g.][and references therein]{BashiZucker2022, Boettner2024, Tsantaki2025, Padois2025, Spitoni2025, Teixeira2025, Webb2026}. Together, these efforts are moving the field towards a genuinely Galactic view of planet formation and evolution.

\section{Conclusions}

Stellar birth radii provide a bridge between Galactic evolution and exoplanet demographics. By combining chemical evolution models with a GAM, we can assign probabilistic formation radii to thin disc stars and compare them with their present-day Galactic orbits. The Solar birth radius provides an intuitive example of why this distinction matters: even the planetary system we know best did not necessarily form at its present Galactic location.

Applied to 1327 planet-hosting stars, this framework indicates that giant-planet systems retain a particularly strong connection to metal-rich inner-disc birth environments, while rocky-only systems are less centrally concentrated. Planet multiplicity shows no clear dependence on radial displacement. Inward-moving hosts extend towards larger \amax, although this trend may partly reflect differences in host-star metallicity rather than a causal effect of Galactic radial displacement.

These results demonstrate the potential of combining exoplanet science with Galactic archaeology. More homogeneous planetary samples, together with improved stellar ages, chemical abundances, and dynamical models, will be required to disentangle the effects of Galactic birth environment, stellar migration, and observational selection.

\acknowledgements
MLLD acknowledges the support of Agencia Nacional de Investigación y Desarrollo (ANID), Chile, through Fondecyt Postdoctorado Folio 3240344. MLLD also thanks ANID Basal Project FB210003. JJGD acknowledges the European Union through the SPARC-UAH Project under Grant SBPLY/23/180225/000071. RS acknowledges support from the National Science Centre, Poland, project 2019/34/E/ST9/00133. MLLD thanks Vardan Adibekyan for a very valuable discussion during IAUS 408 concerning the possible role of host-star metallicity in the orbital-separation trend. MLLD further acknowledges Camilla Danielski for the discussions developed throughout the IAUS 408. AI-assisted tools were used for text review and plot optimisation under author supervision.


\end{document}